\documentclass[onecolumn,showpacs]{revtex4}

\usepackage{graphicx}
\usepackage{dcolumn}
\usepackage{bm}

\input epsf

\begin{document}

\title{\Large  Higher Dimensional Szekeres' Space-time in Brans-Dicke Scalar Tensor Theory}

\author{\bf Asit Banerjee$^1$\footnote{asitb@cal3.vsnl.net.in},~Ujjal Debnath$^2$\footnote{ujjaldebnath@yahoo.com} and~Subenoy
Chakraborty$^2$\footnote{subenoyc@yahoo.co.in}}

\affiliation{$^1$Department of Physics, Jadavpur University,
Calcutta-32, India.\\ $^2$Department of Mathematics, Jadavpur
University, Calcutta-32, India.}

\date{\today}

\begin{abstract}
The generalized Szekeres family of solution for quasi-spherical
space-time of higher dimensions are obtained in the scalar tensor
theory of gravitation. Brans-Dicke field equations expressed in
Dicke's revised units are exhaustively solved for all the
subfamilies of the said family. A particular group of solutions
may also be interpreted as due to the presence of the so-called
C-field of Hoyle and Narlikar and for a chosen sign of the
coupling parameter. The models show either expansion from a big
bang type of singularity or a collapse with the turning point at
a lower bound. There is one particular case which starts from the
big bang, reaches a maximum and collapses with the in course of
time to a crunch.
\end{abstract}


\maketitle

\section{\normalsize\bf{Introduction}}

The first step to solve the Einstein equations for the metric
belonging to Szekeres family[1]
\begin{equation}
ds^{2}=dt^{2}-e^{2\alpha}dr^{2}-e^{2\beta}(dx^{2}+dy^{2})
\end{equation}

was taken by Szekeres [1] with dust and $\Lambda=0$. Here
$\alpha$ and $\beta$ are in general functions of ($t, r, x, y$).
Szekeres' result was generalized by Szafron et al [2] without any
further assumption except for non-zero pressure.  In further
generalizations the perfect fluid has been replaced by a fluid
with heat flow [3], viscosity [4] and also electromagnetic field
[5]. Also Barrow et al [6] gave solutions for dust model with a
cosmological constant and recently Chakraborty et al [7] gave
solutions for perfect fluid model with a cosmological constant in
$(n+2)$-D space-time. We are not aware of any generalization of
the above class in more than 4 dimensions in Brans-Dicke scalar
tensor theory of gravitation. In this paper, we work out
solutions for dust in the presence of the cosmological constant
($\Lambda\ne 0$) and the Brans-Dicke scalar field. We consider
the scalar tensor theory in the Dicke's [8] revised version after
unit transformation. In the revised version, $G$ does not vary
while the masses of the elementary particles are varying. In this
version, the trajectories of particles are not geodesics and the
energy momentum tensor holds for combined matter and the scalar
field. Suitable transformation of units used by Dicke are
$\bar{g}_{\mu\nu}=\lambda g_{\mu\nu}$, $\bar{m}=\lambda^{-1/2}m$,
$d\bar{s}=\lambda^{1/2}ds$ and the scalar field
$\lambda=\phi/\phi_{0}$, where bars indicate the variables in the
revised units, $\lambda$ is the scalar field in the new unit and
$\phi_{0}$ is a constant. The field equations in the revised
units are
\begin{equation}
R_{\mu\nu}-\frac{1}{2}g_{\mu\nu}R=8\pi
G(T_{\mu\nu}+S_{\mu\nu})+\Lambda g_{\mu\nu}
\end{equation}

where ~~~~~$S_{\mu\nu}=\frac{(2\omega+3)}{16\pi
G}\left(C_{\mu}C_{\nu}-\frac{1}{2}g_{\mu\nu}C_{l}C^{l}\right)$
~~ and the scalar~~ $C={\text ln}~\lambda={\text ln}~(\phi/\phi_{0})$.\\

In what follows we consider a more general ($n+2$)-dimensional
Szekeres space-time, which can be expressed by the following
metric form [7]
\begin{equation}
ds^{2}=dt^{2}-e^{2\alpha}dr^{2}-e^{2\beta}\sum^{n}_{i=1}dx_{i}^{2}
\end{equation}

where $\alpha$ and $\beta$ are now functions of all the ($n+2$)
space-time variables i.e., $$\alpha=\alpha(t,r,x_{1},...,x_{n}),~~
\beta=\beta(t,r,x_{1},...,x_{n}).$$

It is to be noted here that in the presence of the non-zero
scalar field it is necessary to make at least one assumption in
order to solve the field equations. We can either assume the form
of scalar field or of the metric co-efficients. Here we have
assumed $\beta'_{x_{i}}=0$. Then there are two possibilities. In
one of these cases the scalar field is in general a function of
$r$ and $t$, whereas in the second case the scalar field may
either be a function of $r$ or a function of $t$ alone. In fact
the differential equation obtained in terms of the metric
co-efficients for $C=C(t,r)$ is apparently not solvable. So in the
next section we present exact solutions in the following
different cases:\\

(i) $\beta'\ne 0, \dot{\beta}_{x_{i}}=0, C=C(t,r)$,\\

(ii) $\beta'\ne 0, \dot{\beta}_{x_{i}}=0, C=C(t)$,\\

(iii) $\beta'= 0, \dot{\beta}_{x_{i}}=0, C=C(t)$.\\

The case $C=C(r)$ is not consistent with the field equations. In
all the above cases the asymptotic behaviour of the dust density
$\rho$ and the shear scalar $\sigma$ are obtained in the limits.
The cosmological models described in all the cases mentioned
above are all inhomogeneous models.\\

The special cases of our solutions in the absence of the scalar
field lead to those of Szekeres as shown in the subsequent
sections. At this point one must note that Szekeres solutions are
however given in co-ordinates different from those used in the
present text.\\

In the last section an alternative interpretation is given for
the above mentioned solutions in the special case of $C$ as a
linear function of time. In this particular case the scalar field
$C$ may be interpreted as the creation field first proposed by
Hoyle and Narlikar [9] in order to obtain a steady state cosmology
or later a quasi steady state cosmology.\\

\section{\normalsize\bf{Field equations in the revised version of the
Brans-Dicke theory and their exact solutions}}

The Einstein's field equations in the presence of the cosmological
constant $\Lambda$ follow from (2) (we choose $8\pi G=1$, see
[7]):

\begin{eqnarray*}
n\dot{\alpha}\dot{\beta}+\frac{1}{2}n(n-1)\dot{\beta}^{2}-e^{-2\beta}\sum_{i=1}^{n}
\left\{\alpha_{x_{i}}^{2}+\frac{1}{2}(n-1)(n-2)\beta_{x_{i}}^{2}+
(n-2)\alpha_{x_{i}}\beta_{x_{i}}+\alpha_{x_{i}x_{i}}\right.
\end{eqnarray*}
\vspace{-8mm}

\begin{equation}
\left.+(n-1)\beta_{x_{i}x_{i}} \right\}+e^{-2\alpha}
\left\{n\alpha'\beta'-\frac{1}{2}n(n+1)\beta'^{2}
-n\beta''\right\}=\Lambda+\rho-\frac{1}{2}f\left(\dot{C}^{2}+e^{-2\alpha}C'^{2}\right)
\end{equation}

\begin{eqnarray*}
\frac{1}{2}n(n+1)\dot{\beta}^{2}+n\ddot{\beta}-\frac{1}{2}n(n-1)e^{-2\alpha}\beta'^{2}
-e^{-2\beta}\sum_{i=1}^{n}\left\{\frac{1}{2}(n-1)(n-2)\beta_{x_{i}}^{2}+
(n-1)\beta_{x_{i}x_{i}}\right\}
\end{eqnarray*}
\vspace{-8mm}

\begin{equation}
=\Lambda+\frac{1}{2}f\left(\dot{C}^{2}+e^{-2\alpha}C'^{2}\right)
\hspace{-3.4in}
\end{equation}

\begin{eqnarray*}
\dot{\alpha}^{2}+\ddot{\alpha}+(n-1)\dot{\alpha}\dot{\beta}+\frac{1}{2}n(n-1)\dot{\beta}^{2}+
(n-1)\ddot{\beta}+e^{-2\alpha}\left\{(n-1)\alpha'\beta'-\frac{1}{2}n(n-1)
\beta'^{2}-\right.
\end{eqnarray*}
\vspace{-8mm}

\begin{eqnarray*}
\left.(n-1)\beta''\right\}-e^{-2\beta}\sum_{i\ne
j=1}^{n}\left\{\alpha_{x_{j}}^{2}+\frac{1}{2}(n-2)(n-3)\beta_{x_{j}}^{2}+
\alpha_{x_{j}x_{j}}+(n-2)\beta_{x_{j}x_{j}}+(n-3)\alpha_{x_{j}}\beta_{x_{j}}\right\}
\end{eqnarray*}
\vspace{-8mm}

\begin{equation}
-e^{-2\beta}\left\{(n-1)\alpha_{x_{i}}\beta_{x_{i}}+
\frac{1}{2}(n-1)(n-2)\beta_{x_{i}}^{2}\right\}=\Lambda+\frac{1}{2}f\left(\dot{C}^{2}
+e^{-2\alpha}C'^{2}\right) \hspace{-.6in}
\end{equation}

\begin{equation}
\alpha_{x_{j}}(-\alpha_{x_{i}}+\beta_{x_{i}})+\beta_{x_{j}}(\alpha_{x_{i}}+
(n-2)\beta_{x_{i}})-\alpha_{x_{i}x_{j}}-(n-2)\beta_{x_{i}x_{j}}=0,~~~
(i\ne j)
\end{equation}

\begin{equation}
\dot{\alpha}\beta'-\dot{\beta}\beta'-\dot{\beta}'=-\frac{1}{n}f\dot{C}C'
\end{equation}

\begin{equation}
-\dot{\alpha}\alpha_{x_{i}}+\dot{\beta}\alpha_{x_{i}}-\dot{\alpha}_{x_{i}}-(n-1)\dot{\beta}_{x_{i}}=0
\end{equation}

\begin{equation}
\alpha_{x_{i}}\beta'-\beta'_{x_{i}}=0
\end{equation}
\\
where dot, dash and subscript stands for partial differentiation
with respect to $t$, $r$ and the corresponding variables
respectively
(e.g., $\beta_{x_{i}}=\frac{\partial\beta}{\partial x_{i}}$) with $i,j=1,2,...,n$ and $f=-\frac{1}{2}(2\omega+3)$.\\

From  equations (8) and (10) after differentiating with respect to
$x_{i}$ and $t$ respectively, we have the integrability condition

\begin{equation}
\dot{\beta}_{x_{i}}\beta'^{2}=-\frac{1}{n}f\dot{C}C'\beta'_{x_{i}},~~~~i=1,2,...,n
\end{equation}\\

This equation cannot be solved without any specific assumption
either on the metric co-efficients or on the scalar field $C$.
Such an assumption is however necessary to obtain exact solutions
for all the variables. One of such assumption is
$\beta'_{x_{i}}=0$, which leave us two possibilities: (i)
$\beta'\ne 0$ so that $\dot{\beta}_{x_{i}}=0$, (ii) $\beta'=0$.
The second choice in view of eq.(8) leads us to the condition
$\dot{C}C'=0$, which implies that $C$ can not be a function of
both the co-ordinates $r$ and $t$. For general $C=C(t,r)$, first
of all we consider $\beta'\ne 0, \dot{\beta}_{x_{i}}=0$. So we
obtain from the field equations (8) and (10), the metric
co-efficients are in the following form:

\begin{equation}
e^{\beta}=R(t,r)~e^{\nu(r,x_{1},...,x_{n})}
\end{equation}
and
\begin{equation}
e^{\alpha}=\frac{R'+R\nu'}{S(t,r)}
\end{equation}

where $R$ and $S$ are function of $t,~r$ only. Now from the field
equations (5) and (6) using equations (12) and (13), we have the
differential equations for $R$ and $S$:

\begin{equation}
2R\ddot{R}+(n-1)(\dot{R}^{2}-b^{2}R^{2})-\frac{1}{n}[2\Lambda+f(\dot{C}^{2}+
e^{-2\alpha}C'^{2})]R^{2}=(n-1)K(r)
\end{equation}
and
\begin{equation}
\frac{f}{2n}~e^{-\alpha}\frac{\partial}{\partial
r}[\dot{C}^{2}+e^{-2\alpha}C'^{2}]=\frac{\dot{S}}{RS^{2}}~\frac{\partial}{\partial
t }(R\dot{S}e^{2\alpha})
\end{equation}

where $K$ is a function of $r$ only. Since $\beta'_{x_{i}}=0$, so
$\alpha$ is a function of $t,~r$ only and the form of $e^{\nu}$
suggests

\begin{equation}
e^{-\nu}=A(r)\left(\sum_{i=1}^{n}(x_{i}^{2}+x_{i})+1\right)
\end{equation}

where $A(r)$ is arbitrary function of $r$ alone satisfying
~$(n-4)A^{2}(r)=K(r)$. It is to be noted that when $n\ne 4$ then
$K(r)\ne 0$ and $n=4$ implies $K(r)=0$, which shows that $K(r)$
must vanish when we consider a six dimensional space-time. In
other cases, however $K(r)\ne 0$. The field equation (7) is
automatically satisfies by the above solutions. Also from the
field equation (4), the expression for the energy density $\rho$
is

\begin{equation}
\rho(t,r,x_{1},...,x_{n})=-\frac{n}{n-1}(\ddot{\alpha}+n\ddot{\beta}+\dot{\alpha}^{2}
+n\dot{\beta}^{2})+\frac{2\Lambda}{n-1}+\frac{nf}{n-1}(\dot{C}^{2}+e^{-2\alpha}C'^{2})
\end{equation}

and the expression for the shear scalar is

\begin{equation}
\sigma^{2}=\frac{n}{2(n+1)}\frac{(R\dot{R}'-\dot{R}R')^{2}}{R^{2}(R'+R\nu')^{2}}
\end{equation}

Now eliminating the terms containing the derivative of the scalar
field $C$ between (14) and (15) we obtain the differential
equation in $R$ and $S$ as

\begin{equation}
\frac{\partial}{\partial
r}\left[\frac{1}{R^{2}}\left\{2R\ddot{R}+(n-1)(\dot{R}^{2}-S^{2}-K)\right\}
\right]=2e^{\alpha}\frac{\dot{S}}{RS^{2}}\frac{\partial}{\partial
t}(R\dot{S}e^{2\alpha})
\end{equation}

This differential equation in $R$ and $S$ is quite complicated
and is apparently not solvable in closed form. For general
$C=C(t,r)$, it is not possible to find out any exact solutions.
So we may consider $C$ is a function of $t$ or $r$ alone. For
$C=C(t)$, we arrive the following two cases from equation (11) in
order to obtain explicit form of the metric co-efficients: (i)
$\beta'\ne 0,~ \dot{\beta}_{x_{i}}=0$, (ii) $\beta'=0$.\\\\

{\bf Case I}:~~ $\beta'\ne 0,~ \dot{\beta}_{x_{i}}=0 ~(i=1,2,...,n),~ C=C(t):$\\

With the above condition being used we obtain from the field
equations (8) and (10), the metric coefficients in the following
form [7]:

\begin{equation}
e^{\beta}=R(t,r)~e^{\nu(r,x_{1},...,x_{n})}
\end{equation}

and
\begin{equation}
e^{\alpha}=R'+R~\nu'
\end{equation}

Now using the relation (20) and (21) in equation (5), we obtain a
differential equation for $R$ as follows

\begin{equation}
2R\ddot{R}+(n-1)\dot{R}^{2}-\frac{1}{n}(2\Lambda+f\dot{C}^{2})R^{2}=(n-1)K
\end{equation}

and the solution for $\nu$ can be expressed as

\begin{equation}
e^{-\nu}=A\sum_{i=1}^{n}x_{i}^{2}+\sum_{i=1}^{n}B_{i}x_{i}+D~.
\end{equation}

There is, however, the following restriction

\begin{equation}
\sum_{i=1}^{n}B_{i}^{2}-4AD=K-1
\end{equation}

where $K,~A,~B_{i}$'s and $D$ are all functions of the
co-ordinate `$r$' alone.\\

The first integral of the equation (22) may be given by

\begin{equation}
\dot{R}^{2}=K+\frac{F}{R^{n-1}}+\frac{2\Lambda}{n(n+1)}R^{2}
+\frac{1}{n}~fR^{1-n}\int \dot{C}^{2}R^{n}\dot{R}dt
\end{equation}

where $F$ is an arbitrary functions of `$r$' and $f$ is a
constant which is already expressed earlier in terms of
Brans-Dicke parameter $\omega$.\\

The following situations are considered separately:\\

(i)~ $C=C_{0}t,~~K\ne 0$ and $n=3$, that is this satisfied
$\lambda=e^{C_{0}t}$ where $C_{0}$ is a constant.\\

The integration of the equation (25) in the next step yields

\begin{equation}
2zR^{2}+K=(4zF-K^{2})^{1/2}~Sinh\{2z^{1/2}(t-t_{0})\},~~z>0
\end{equation}
and
\begin{equation}
2|z|R^{2}-K=(K^{2}+4|z|F)^{1/2}~Sin\{2|z|^{1/2}(t-t_{0})\},~~z<0
\end{equation}

In the above solution for $R$ is another arbitrary function of
$r$ and $z$ is a constant given by
$z=\frac{(2\Lambda+fC_{0}^{2})}{n(n+1)}$.\\

(ii)~ $C=C_{0}t,~ K=0$, ~$n$ is arbitrary:\\

The solution is given by

\begin{equation}
R^{\frac{n+1}{2}}=\left(\frac{F}{z}\right)^{1/2}Sinh\{\frac{1}{2}(n+1)z^{1/2}(t-t_{0})\},~~z>0
\end{equation}
and
\begin{equation}
R^{\frac{n+1}{2}}=\left(\frac{F}{|z|}\right)^{1/2}Sin\{\frac{1}{2}(n+1)|z|^{1/2}(t-t_{0})\},~~z<0
\end{equation}

Here of course there is another possibility. If $F<0$, we can
obtain from (25) on integration a different solution expressed as

\begin{equation}
R^{\frac{n+1}{2}}=\left(\frac{|F|}{z}\right)^{1/2}Cosh\{\frac{1}{2}(n+1)z^{1/2}(t-t_{0})\},~~z>0
\end{equation}

The solutions (28) and (30) have different properties and will be
discussed subsequently.\\

(iii)~ $C=C_{0}t^{a},~K=0$, ~`$a$' being a constant parameter. We
put $R=G^{2/(n+1)}$ in the equation (22) so that we obtain

$$
4nS^{-1}\ddot{G}-(n+1)(2\Lambda+f\dot{C}^{2})=0
$$

which on assume $\Lambda=0$ may also be written as

\begin{equation}
\ddot{G}-yt^{2a-2}G=0
\end{equation}

where $y=\frac{(n+1)}{4n}fC_{0}^{2}a^{2}$. In equation (31) after
integration finally yields the following solution for the
variable $R$,

\begin{equation}
G\equiv
R^{\frac{n+1}{2}}=t^{1/2}\left[A_{1}(r)~I_{\frac{1}{2a}}[y^{1/2}t^{a}/a]+
A_{2}(r)~I_{-\frac{1}{2a}}[y^{1/2}t^{a}/a]\right],~~y>0
\end{equation}
and
\begin{equation}
G\equiv
R^{\frac{n+1}{2}}=t^{1/2}\left[A_{1}(r)~J_{\frac{1}{2a}}[|y|^{1/2}t^{a}/a]+
A_{2}(r)~J_{-\frac{1}{2a}}[|y|^{1/2}t^{a}/a]\right],~~y<0
\end{equation}

It may be noted that when $a=1$, the solutions (32) reduces to
either (28) or (30) but solution (33) reduces to (29) only. As
usual $J_{m}(x)$ and $I_{m}(x)$ respectively stand for Bessel
function and modified Bessel function of first kind of order $m$.\\

The expressions for the density $\rho$ and the shear scalar
$\sigma$ are obtained from (17) and (18) and are given by

\begin{equation}
\rho(t,r,x_{1},...,x_{n})=-\frac{n}{(n-1)}\left[\frac{\ddot{R}'+\ddot{R}\nu'}
{R'+R\nu'}+\frac{n\ddot{R}}{R}\right]+\frac{(2\Lambda+nf\dot{C}^{2})}{(n-1)}
\end{equation}

and

\begin{equation}
\sigma^{2}=\frac{n}{2(n+1)}\frac{(R\dot{R}'-\dot{R}R')^{2}}{R^{2}(R'+R\nu')^{2}}
\end{equation}

The explicit forms of (34) and (35) are apparently very
complicated. However one can discuss the behaviour of different
models in the limits $t\rightarrow t_{0}$ and $t\rightarrow
\infty$. In fact $t_{0}(r)$ depends the initial moment of
evolution in each case. The evolution may or may not begin with a
big bang singularity. Again since $t_{0}'\ne 0$ the instant of
singularity if it exists will be position dependent. These are
apparent from the following facts.\\

In case (i) with $K\ne 0$ the dust density and the shear scalar
both are finite at $t\rightarrow t_{0}$ but in the other extreme
limit that is as $t\rightarrow \infty$ the dust density $\rho$
remains finite but the shear scalar $\sigma$ vanishes.\\

In case (ii), $\rho\rightarrow \infty$ and $\sigma\rightarrow
\infty$ at the initial instant $t\rightarrow t_{0}$ but in course
of time as $t\rightarrow \infty$ the dust density remains finite,
whereas $\sigma\rightarrow 0$. This is true for the solution (28).
On the other hand if we concentrate our attention on the solution
(30). We find that the density $\rho$ remains finite throughout
the evolution, whereas the shear starts from a finite
magnitude and gradually disappear in course of evolution.\\

In case (iii), we note that there is singularity ($R=0$) for the
solution in equations (32) and (33) provided we choose arbitrary
constants $A_{2}(r)=0$. Also asymptotically for large $t$, $R$
oscillates infinitely for the solution (33) while $R$ becomes
infinite for the solution (32). In this case, for $0 \le a \le 1$,
$\rho$ and $\sigma$ both explode initially whereas they attain
finite magnitudes in course of evolution as $t\rightarrow \infty$.\\\\

{\bf Case II}:~~ $\beta'= 0,~ C=C(t)$:\\

In this case, for only the above choices, the exact solutions can
not be found from the field equations (that shown by Szekeres
[1]). So we need further assumption like $\dot{\beta}_{x_{i}}=0~
(i=1,2,...,n)$,
in order to obtain the exact solutions.\\

Now from the field equations we obtain the metric coefficients in
the form

\begin{equation}
e^{\beta}=R(t)~e^{\nu(x_{1},x_{2},...,x_{n})}
\end{equation}

and
\begin{equation}
e^{\alpha}=R(t)~\eta(r,x_{1},x_{2},...,x_{n})+\mu(t,r)
\end{equation}

Then as before from the field equation (5), we have similar
differential equations in $R$ as

\begin{equation}
2R\ddot{R}+(n-1)\dot{R}^{2}-\frac{1}{n}(2\Lambda+f\dot{C}^{2})R^{2}=(n-1)K
\end{equation}

and the solution for $\nu$ as

\begin{equation}
e^{-\nu}=A\sum_{i=1}^{n}x_{i}^{2}+\sum_{i=1}^{n}B_{i}x_{i}+D
\end{equation}

along with the restriction

\begin{equation}
\sum_{i=1}^{n}B_{i}^{2}-4AD=K
\end{equation}

Here $K,~A,~B_{i}$'s and $D$ are all arbitrary constants.\\

Now from the field equation (7) we have the solution for $\eta$ as

\begin{equation}
e^{-\nu}\eta=u\sum_{i=1}^{n}x_{i}^{2}+\sum_{i=1}^{n}v_{i}x_{i}+w
\end{equation}

and the resulting differential equation in $\mu$ is

\begin{equation}
R\ddot{\mu}+(n-1)\dot{R}\dot{\mu}+\mu\left[\ddot{R}-\frac{1}{n}~(2\Lambda+
f\dot{C}^{2})R\right]=g(r)
\end{equation}

with
\begin{equation}
g(r)=(n-1)\left[2(uD+wA)-\sum_{i=1}^{n}v_{i}B_{i}\right]
\end{equation}

where $u,~ v_{i}$'s and $w$ as arbitrary functions of the co-ordinate `r' alone.\\

For simplicity, let us choose $C=C_{0}t,~ n=3$. In this case the
solutions of $R$ (as before) and $\mu$ are

\begin{equation}
2zR^{2}+K=(4zF-K^{2})^{1/2}~Sinh\{2z^{1/2}(t-t_{0})\},~~z>0,
\end{equation}
\begin{equation}
2|z|R^{2}-K=(K^{2}-4zF)^{1/2}~Sin\{2|z|^{1/2}(t-t_{0})\},~~z<0
\end{equation}

and
\begin{equation}
2z\mu
R+g(r)=\sqrt{4z^{2}h(r)-g^{2}(r)}~Sinh\left\{\sqrt{2z}~(t-t_{1}(r))\right\},~~z>0,
\end{equation}
\begin{equation}
2|z|\mu
R-g(r)=\sqrt{g^{2}(r)-4z^{2}h(r)}~Sin\left\{\sqrt{2|z|}~(t-t_{1}(r))\right\},~~z<0,
\end{equation}

where $t_{0},~F$ are arbitrary constants, $h(r),~t_{1}(r)$ are
arbitrary functions of `$r$' only and
$z=\frac{1}{12}(2\Lambda+fC_{0}^{2})$. The solutions (44) and (45)
is identical with (28) and (29) except for the fact that now $F$
and $t_{0}$ are no longer functions of $r$.\\

In this case the expressions for the density and the shear scalar
are given by

\begin{equation}
\rho(t,r,x_{1},...,x_{n})=-\frac{n}{(n-1)}\left[\frac{\ddot{\mu}+\ddot{R}\eta}
{\mu+R\eta}+\frac{n\ddot{R}}{R}\right]+\frac{(2\Lambda+nf\dot{C}^{2})}{(n-1)}
\end{equation}

\begin{equation}
\sigma^{2}=\frac{n}{2(n+1)}\frac{(R\dot{\mu}-\dot{R}\mu)^{2}}{R^{2}(\mu+R\eta)^{2}}
\end{equation}

There are two distinct cases for either $K\ne 0$ or $K=0$. In the
former case with $K<0$ both the dust density and the shear scalar
remain finite when $t\rightarrow t_{0}$, but they are infinitely
large at this instant in the second case $K=0$. On the other hand
as $t\rightarrow \infty$ the density is finite even though the
shear vanishes in course of evolution. When $K>0$ the situation
differs. Here the density and shear explode at same initial
instant other than $t=t_{0}$.\\\\

Next if we consider $C=C(r)$ then from equation (11), we may get
also two possibilities: (i) $\beta'\ne 0,~
\dot{\beta}_{x_{i}}=0$, (ii) $\beta'=0$. In these cases the form
of metric co-efficients are similar to case I and Case II, but
the differential equations (22) and (38) are slightly different
i.e.,
$$
2R\ddot{R}+(n-1)\dot{R}^{2}-\frac{1}{n}(2\Lambda+fe^{-2\alpha}C'^{2})R^{2}=(n-1)K
$$

This equation can not be solved because $\alpha$ is a function of
all space-time co-ordinates. So we are not going to further
discussion on the choice $C=C(r)$.\\

\section{\normalsize\bf{Alternative interpretation of the solutions in terms of
$C$-field cosmology}}

It is appropriate at this stage to interpret some of the
previously described solutions as due to the presence of the
creation field first introduced by Hoyle and Narlikar [9]. The
field equations in this case are exactly identical with (2)
except for the replacement of $S_{\mu\nu}$ by $T^{(c)}_{\mu\nu}$,
where
\begin{equation}
T^{(c)}_{\mu\nu}=-f\left(C_{\mu}C_{\nu}-\frac{1}{2}~g_{\mu\nu}C_{l}C^{l}\right)
\end{equation}

where $C_{\mu}=C,_{\mu}$ and $f~(>0)$ is a coupling constant. The
additional feature in $C$-field cosmology is that one must confirm
that the $C$-field satisfies the source equation
\begin{equation}
fC_{;~\mu}^{\mu}=j_{;~\mu}^{\mu}
\end{equation}

with $j^{\mu}=\rho\frac{dx^{\mu}}{ds}$.\\

It is not difficult to check that the Bianchi identity and the
source equation (51) lead to the relation $\dot{C}=1$, which
determines the expression of the $C$-field scalar $C=t+\psi(r)$.
So for obvious reasons all the previous solutions with the scalar
field~ $C=\text{ln}~\lambda$~ expressed as a linear function of
time are also solutions of the $C$-field cosmology.

\section{\normalsize\bf{Discussions}}

One must note that in all the models presents above some are
singularity free and some have big bang type singularities at the
starting point. Particularly for the function $K=0$ those which
show the turning point ($\dot{R}=0$) at some stage represent only
the lower bound (since $\ddot{R}>0$) as are evident from from the
equations (22) and (38). The singularity occurs in each case at
an instant which is not fixed for different shells of different
$r$. In fact, the singularity $R=0$ is position dependent in all
the cases described above and the reason is that the system is
inhomogeneous which is the limit of spherical dust reduces to
Tolman-Bondi space-time. There is one particular case given by
the solution (29) where there is an initial singularity at
$t=t_{0}$. Subsequently $R(t,r)$ increases
and reaches a maximum followed by a collapse to a crunch.\\

Another interesting point to observe is that for a few solutions
such as (26), (28), (44) etc, we have $\ddot{R}$ initially less
than zero but becomes positive in course of time indicating that
the expansion starts from decelerating phase to an accelerating
phase at late stage. It is to be noted that in the above set of
solutions is a kind of singularity given by $e^{\alpha}=0$ or
equivalently $\left(e^{\beta}\right)'=0$, which are analogous to
the shell crossing singularity in Tolman-Bondi
[10] models. In this case also the density $\rho$ diverges. Finally
it can be mentioned that one can obtain Szekeres 4 dimensional solutions
from our solutions if we put $n=2, \Lambda=0$ and $C=0$. In fact the differential
equations in $R$ (see equations (22) and (38)) after the above simplification
becomes identical to the corresponding differential equations in Szekeres
solutions [1]. Therefore our solutions are generalization of Szekeres solutions.\\

{\bf Acknowledgement:}\\

One of the authors (U.D) is thankful to CSIR (Govt. of India)
for awarding a Senior Research Fellowship. Authors are also
thankful to the referee for his valuable comments which help
to improve the paper.\\

{\bf References:}\\
\\
$[1]$  P. Szekeres, {\it Commun. Math. Phys.} {\bf 41} 55 (1975); A. Krasinski,
{\it Inhomogeneous Cosmological Models} (Cambridge Univ. Press, Cambridge England, 1997).\\
$[2]$  D. A. Szafron, {\it J. Math. Phys.} {\bf 18} 1673 (1977);
D. A. Szafron and J. Wainwright {\it J. Math. Phys.}
{\bf 18} 1668 (1977).\\
$[3]$ S. W. Goode, {\it Class. Quantum Grav.} {\bf 3} 1247 (1986).\\
$[4]$ S. R. Roy, J. P. Singh, {\it Indian. J. Pure Appl. Math.}
{\bf 13} 1285 (1982).\\
$[5]$ K. A. Bronnikov {\it Gen. Rel. Grav.} {\bf 15} 823 (1983).\\
$[6]$  J. D. Barrow and J. Stein-Schabes, {\it Phys. Letts.} {\bf 103A} 315 (1984).\\
$[7]$ S. Chakraborty and U. Debnath, gr-qc/0304072.\\
$[8]$ R. H. Dicke, {\it Phys. Rev. D} {\bf 125} 2163 (1962).\\
$[9]$ F. Hoyle and J. V. Narlikar, {\it Proc. R. Soc. A} {\bf
273} 1 (1963); {\it Proc. R. Soc. A} {\bf 278} 465 (1964).\\
$[10]$ A. Banerjee, A. Sil and S. Chatterjee, {\it Astrophys. J.}
{\bf 422} 681 (1994); C. Hellaby and K. Lake, {\it Astrophys. J.}
{\bf 290} 381 (1985);
C. Hellaby and A. Krasinski, {\it Phys. Rev. D} {\bf 66} 084011 (2002).\\

\end{document}